\documentclass[twocolumn,superscriptaddress,floatfix,showpacs]{revtex4-1}
\usepackage{amsmath,amsfonts,amssymb}
\usepackage[pdftex]{color,graphicx,hyperref}
\pdfoutput=1
\usepackage{subfigure}
\usepackage{graphicx}
\usepackage{multirow}
\usepackage{array}
\usepackage{colortbl}

\begin{document}

\title{Universal features of correlated bursty behaviour}
\author{M. Karsai} 
\email{marton.karsai@aalto.fi} 
\author{K. Kaski}
\affiliation{BECS, School of Science, Aalto University, P.O. Box 12200, FI-00076}
\author{A.-L. Barab\'asi}
\affiliation{Center for Complex Networks Research, Northeastern University, Boston, MA 02115}
\affiliation{Institute of Physics and BME-HAS Cond. Mat. Group, BME, Budapest, Budafoki \'ut 8., H-1111}
\author{J. Kert\'esz}
\affiliation{Institute of Physics and BME-HAS Cond. Mat. Group, BME, Budapest, Budafoki \'ut 8., H-1111}
\affiliation{BECS, School of Science, Aalto University, P.O. Box 12200, FI-00076}
\date{\today}

\begin{abstract}
Inhomogeneous temporal processes, like those appearing in human communications, neuron spike trains, and seismic signals, consist of
high-activity bursty intervals alternating with long low-activity periods. In recent studies such bursty behavior has been characterized
by a fat-tailed inter-event time distribution, while temporal correlations were measured by the autocorrelation function. However,
these characteristic functions are not capable to fully characterize temporally correlated heterogenous behavior. Here we show that the
distribution of the number of events in a bursty period serves as a good indicator of the dependencies, leading to the universal
observation of power-law distribution in a broad class of phenomena. We find that the correlations in these quite different systems can be
commonly interpreted by memory effects and described by a simple phenomenological model, which displays temporal behavior qualitatively
similar to that in real systems.
\end{abstract}

\maketitle

\begin{figure}[h!] \centering
\includegraphics[width=80mm]{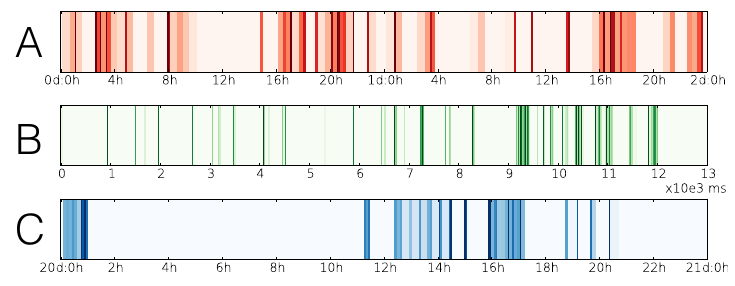}
\caption{Activity of single entities with color-coded inter-event times. (a): Sequence of earthquakes at a single location (b): Firing sequence of a single neuron (c): Outgoing mobile phone call sequence of an individual. Shorter the time between the consecutive events darker the color.}
\label{fig:shem1}
\end{figure}

In nature there are various phenomena, from earthquakes \cite{Corral1} to sunspots \cite{Wheatland1} and neuronal activity, \cite{Kemuriyama1} that show temporally inhomogeneous sequence of events, in which the overall dynamics is determined by aggregate effects of competing processes. This happens also in human dynamics as a result of individual decision making and of various kinds of correlations with one's social environment. These systems can be characterized by intermittent switching between periods of low activity and high activity bursts \cite{Barabasi1,Oliveira1,BarabasiBursts}, which can appear as a collective phenomena similar to processes seen in self-organized criticality \cite{Bak1,Bak2, Jensen1,Paczuski1,Zapperi1,Beggs1,Lippiello1,Arcangelis1,Brunk1,Ramos1}. In contrast with such self-organized patterns intermittent switching can be detected at the individual level as well (see Fig.\ref{fig:shem1}), as seen for single neuron firings and for earthquakes at a single location, both showing inhomogeneous temporal patterns \cite{Kepecs1,Grace1,Smalley1,Udias1,Zhao1,Vazquez1}.

Further examples of bursty behavior at the individual level have been observed in the digital records of human communication activities through different channels \cite{Barabasi1,Goh1,Eckmann1,Ciamarra1,Ratkiewicz1}. Over the last few years different explanations have been proposed about the origin of inhomogeneous human dynamics \cite{Barabasi1,Vazquez1,Malmgren1}, including the single event level \cite{Wu1}, and about the impact of circadian and weekly fluctuations \cite{Jo1}. Moreover, by using novel technology of Radio Frequency ID's, heterogeneous temporal behavior was observed in the dynamics of face-to-face interactions \cite{Cattuto:2010,Takaguchi1}. This was explained by a reinforcement dynamics \cite{Stehle:2010,Zhao:2011} driving the decision making process at the single entity level.

For systems with discrete event dynamics it is usual to characterize the observed temporal inhomogeneities by the inter-event time distributions, $P(t_{ie})$, where $t_{ie}=t_{i+1}-t_{i}$ denotes the time between consecutive events. A broad $P(t_{ie})$ \cite{Kemuriyama1,Goh1,Saichev1} reflects large variability in the inter-event times and denotes heterogeneous temporal behavior. Note that $P(t_{ie})$ alone tells nothing about the presence of correlations, usually characterized by the autocorrelation function,  $A(\tau)$, or by the power spectrum density. However, for temporally heterogeneous signals of independent events with fat-tailed $P(t_{ie})$ the Hurst exponent can assign false positive correlations \cite{Hansen1} together with the autocorrelation function (see Supplementary Information). To understand the mechanisms behind these phenomena, it is important to know whether there are true correlations in these systems. Hence for systems showing fat-tailed inter-event time distributions, there is a need to develop new measures that are sensitive to correlations but insensitive to fat tails.

In this paper we define a new measure that is capable of detecting whether temporal correlations are present, even in the case of heterogeneous signals. By analyzing the empirical datasets of human communication, earthquake activity, and neuron spike trains, we observe universal features induced by temporal correlations. In the analysis we establish a close relationship between the observed correlations and memory effects and propose a phenomenological model that implements memory driven correlated behavior.

\section{Results}

\subsection{Correlated events}

A sequence of discrete temporal events can be interpreted as a time-dependent point process, $X(t)$, where $X(t_i)=1$ at each time step $t_i$ when an event takes place, otherwise $X(t_i)=0$. To detect bursty clusters in this binary event sequence we have to identify those events we consider correlated. The smallest temporal scale at which correlations can emerge in the dynamics is between consecutive events. If only $X(t)$ is known, we can assume two consecutive actions at $t_i$ and $t_{i+1}$ to be related if they follow each other within a short time interval, $t_{i+1}-t_i\leq \Delta t$  \cite{Wu1,Turnbull1}. For events with the duration $d_i$ this condition is slightly modified: $t_{i+1}-(t_i+d_i) \leq \Delta t$.

\begin{figure*}[ht!] \centering
\includegraphics[width=160mm]{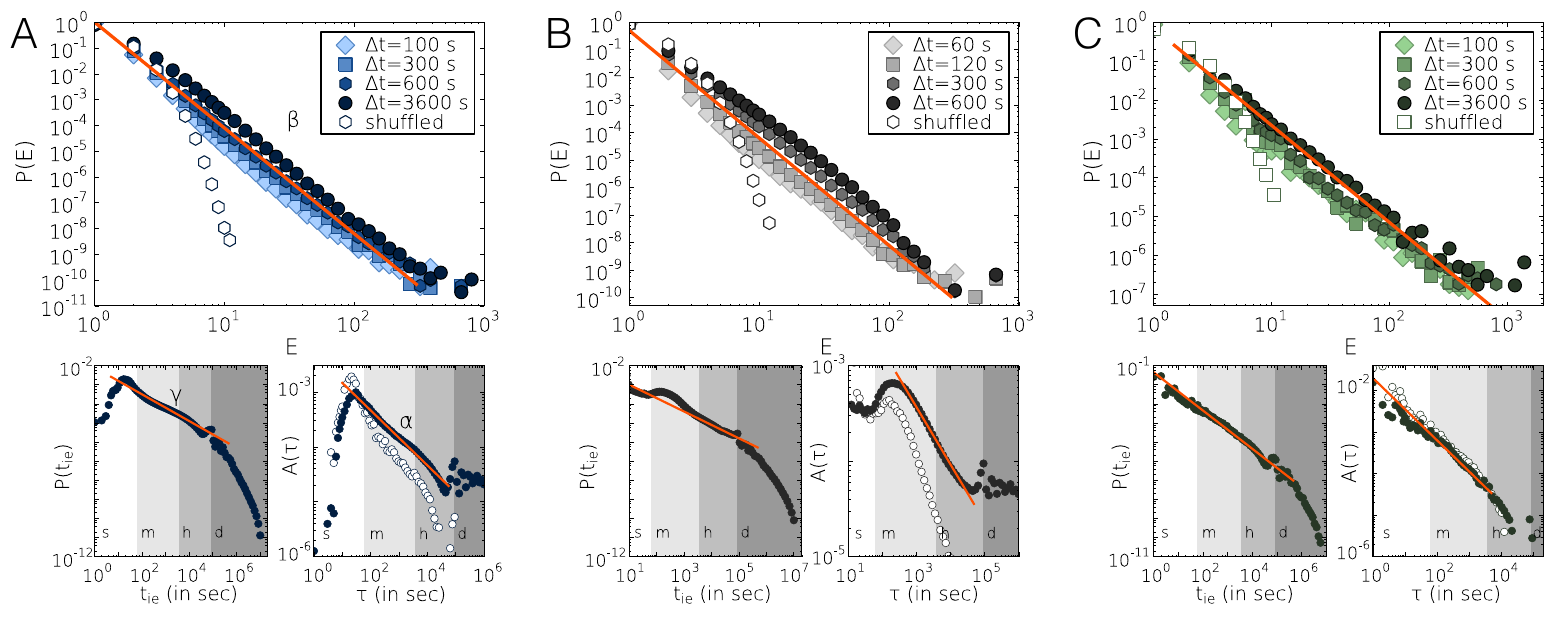}
\caption{The characteristic functions of human communication event sequences. The $P(E)$ distributions with various $\Delta t$ time-window sizes (main panels), $P(t_{ie})$ distributions (left bottom panels) and average autocorrelation functions (right bottom panels) calculated for different communication datasets. (a) Mobile-call dataset: the  scale-invariant behavior was characterized by power-law functions with exponent values $\alpha\simeq 0.5$, $\beta\simeq 4.1$ and  $\gamma\simeq 0.7$ (b) Almost the same exponents were estimated for short message sequences taking values $\alpha\simeq 0.6$, $\beta\simeq 3.9$ and  $\gamma\simeq 0.7$. (c) Email event sequence with estimated exponents $\alpha\simeq 0.75$, $\beta\simeq 2.5$ and $\gamma=1.0$. Empty symbols assign the corresponding calculation results on independent sequences. Lanes labeled with s, m, h and d are denoting seconds, minutes, hours and days respectively.}
\label{fig:1}
\end{figure*}

This definition allows us to detect bursty periods, defined as a sequence of events where each event follows the previous one within a time interval $\Delta t$. By counting the number of events, $E$, that belong to the same bursty period, we can calculate their distribution $P(E)$ in a signal. For a sequence of independent events, $P(E)$ is uniquely determined by the inter-event time distribution $P(t_{ie})$ as follows:
\begin{equation}
P(E=n)= \left(  \int_0^{\Delta t}P(t_{ie})dt_{ie} \right)^{n-1} \left( 1-\int_0^{\Delta t}P(t_{ie})dt_{ie} \right)
\label{eq:0}
\end{equation}
for $n>0$. If the measured time window is finite, the integral $\int_0^{\Delta t}P(t_{ie})dt_{ie}<1$ and $P(E=n)\sim a^{-(n-1)}$, otherwise if $\int_0^{\Delta t}P(t_{ie})dt_{ie}=1$ then $P(E=n)=0$ (for related numerical results see SI). Consequently for any finite independent event sequence the $P(E)$ distribution decays exponentially even if the inter-event time distribution is fat-tailed. Deviations from this exponential behavior indicate correlations in the timing of the consecutive events.

\subsubsection{Bursty sequences in human communication}

To check the scaling behavior of $P(E)$ in real systems we focused on outgoing events of individuals in three selected datasets: (a) A mobile-call dataset from a European operator; (b) Text message records from the same dataset; (c) Email communication sequences \cite{Eckmann1} (for detailed data description see Materials and Methods). For each of these event sequences the distribution of inter-event times measured between outgoing events are shown in Fig.\ref{fig:1} (left bottom panels) and the estimated power-law exponent values are summarized in Table \ref{table:1}. To explore the scaling behavior of the autocorrelation function, we took the averages over $1,000$ randomly selected users with maximum time lag of $\tau= 10^6$. In Fig.\ref{fig:1}.a and b (right bottom panels) for mobile communication sequences strong temporal correlation can be observed (for exponents see Table \ref{table:1}). The power-law behavior in $A(\tau)$ appears after a short period denoting the reaction time through the corresponding channel and lasts up to $12$ hours, capturing the natural rhythm of human activities. For emails in Fig.\ref{fig:1}.c (right bottom panels) long term correlation are detected up to $8$ hours, which reflects a typical office hour rhythm (note that the dataset includes internal email communication of a university staff).

\begin{table}[ht!]
\begin{center}
\begin{tabular}{ p{4.3cm} p{.8cm} p{.8cm} p{.8cm} p{.8cm}}
\rowcolor[gray]{.8}[1\tabcolsep] \hline
  & $\alpha$ & $\beta$ & $\gamma$ & $\nu$\\ \hline
 Mobile-call sequence & $0.5$ & $4.1$ & $0.7$ & $3.0$\\ \hline
\rowcolor[gray]{.8}[1\tabcolsep] 
 Short message sequence & $0.6$ & $3.9$ & $0.7$ & $2.8$\\ \hline
 Email sequence& $0.75$ & $2.5$ & $1.0$ & $1.3$\\ \hline
\rowcolor[gray]{.8}[1\tabcolsep]
 Earthquake sequence (Japan) & $0.3$ & $2.5$ & $0.7$ & $1.6$\\ \hline
 Neuron firing sequence & $1.0$ & $2.3$ & $1.1$ & $1.3$\\ \hline
 \rowcolor[gray]{.8}[1\tabcolsep]
 Model & $0.7$ & $3.0$ & $1.3$ & $2.0$\\ \hline
\end{tabular} 
\end{center}
\caption{Characteristic exponents of the ($\alpha$) autocorrelation function, ($\beta$) bursty number, ($\gamma$) inter-event time distribution functions and $\nu$ memory functions (see SI) calculated in different datasets and for the model study.}
\label{table:1}
\end{table}

The broad shape of $P(t_{ie})$ and $A(\tau)$ functions confirm that human communication dynamics is inhomogeneous and displays non-trivial correlations up to finite time scales. However, after destroying event-event correlations by shuffling inter-event times in the sequences (see Materials and Methods) the autocorrelation functions still show slow power-law like decay (empty symbols on bottom right panels),  indicating spurious unexpected dependencies. This clearly demonstrates the disability of $A(\tau)$ to characterize correlations for heterogeneous signals (for further results see SI). However, a more effective measure of such correlations is provided by $P(E)$. Calculating this distribution for various $\Delta t$ windows, we find that the $P(E)$ shows the following scale invariant behavior
\begin{equation}
P(E)\sim E^{-\beta}
\label{eq:E}
\end{equation} 
for each of the event sequences  as depicted in the main panels of Fig.\ref{fig:1}. Consequently $P(E)$ captures strong temporal correlations in the empirical sequences and it is remarkably different from $P(E)$ calculated for independent events, which, as predicted by (\ref{eq:0}), show exponential decay (empty symbols on the main panels). 

Exponential behavior of $P(E)$ was also expected from results published in the literature assuming human communication behavior to be uncorrelated \cite{Malmgren1,Wu1,Anteneodo1}. However, the observed scaling behavior of $P(E)$ offers direct evidence of correlations in human dynamics, which can be responsible for the heterogeneous temporal behavior. These correlations induce long bursty trains in the event sequence rather than short bursts of independent events.

We have found that the scaling of the $P(E)$ distribution is quite robust against changes in $\Delta t$ for an extended regime of time-window sizes (Fig.\ref{fig:1}). In addition, the measurements performed on the mobile-call sequences indicate that the $P(E)$ distribution remains fat-tailed also when it is calculated for users grouped by their activity. Moreover, the observed scaling behavior of the characteristic functions remains similar if we remove daily fluctuations (for results see SI). These analyses together show that the detected correlated behavior is not an artifact of the averaging method nor can be attributed to variations in activity levels or circadian fluctuations.

\begin{figure}[ht!] \centering
\includegraphics[width=70mm]{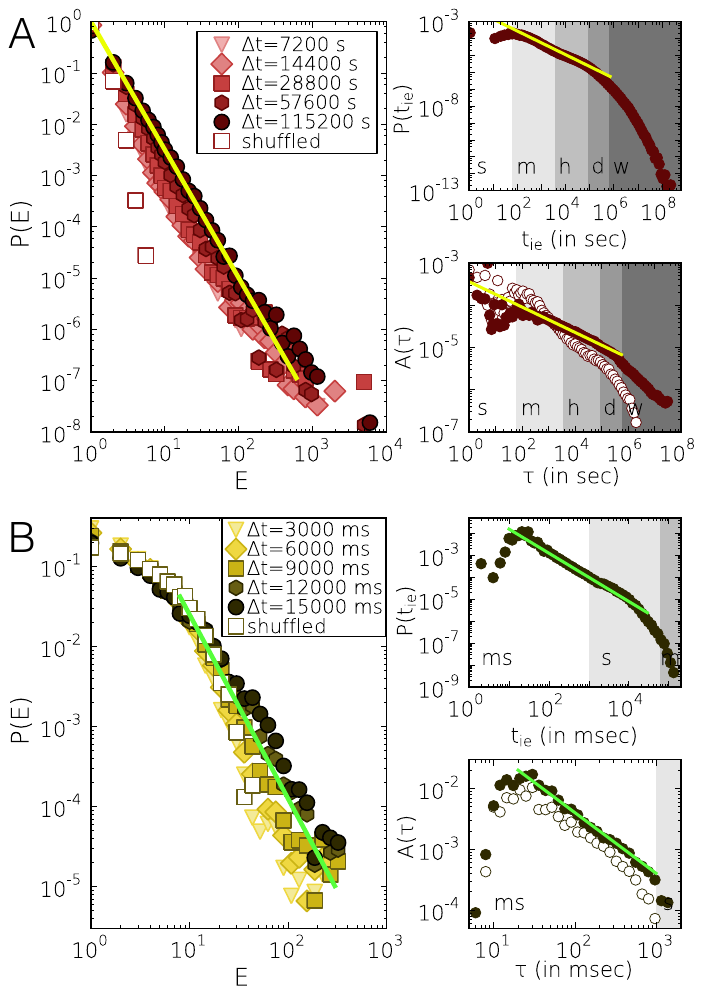}
\caption{The characteristic functions of event sequences of natural phenomena. The $P(E)$ distributions of correlated event numbers with various $\Delta t$ time-window sizes (main panels), $P(t_{ie})$ distributions (right top panels) and average autocorrelation functions (right bottom panels). (a) One station records of Japanese earthquake sequences from 1986 to 1998. The functional behavior is characterized by the fitted power-law functions with corresponding exponents $\alpha\simeq 0.3$, $\beta\simeq 2.5$ and $\gamma\simeq 0.7$. Inter-event times for $P(t_{ie})$ were counted with $10$ second resolution. (b) Firing sequences of single neurons with $2$ millisecond resolution. The corresponding exponents take values as $\alpha\simeq 1.0$, $\beta\simeq 2.3$ and $\gamma\simeq 1.1$. Empty symbols assign the calculation results on independent sequences. Lanes labeled with ms, s, m, h, d and w are denoting milliseconds, seconds, minutes, hours, days and weeks respectively.}
\label{fig:2}
\end{figure}

\subsubsection{Bursty periods in natural phenomena}

As discussed above, temporal inhomogeneities are present in the dynamics of several natural phenomena, e.g. in recurrent seismic activities at the same location \cite{Smalley1,Udias1,Zhao1} (for details see Materials and Methods and SI). The broad distribution of inter-earthquake times in Fig.\ref{fig:2}.a (right top panel) demonstrates the temporal inhomogeneities and the long tail of the autocorrelation function (right bottom panel) assigning long-range temporal correlations. Counting the number of earthquakes belonging to the same bursty period with $\Delta t=2...32$ hours window sizes, we obtain a broad $P(E)$ distribution (see Fig.\ref{fig:2}.a main panel), as observed earlier in communication sequences, but with a different exponent value (see in Table \ref{table:1}). Note that the presence of long bursty trains in earthquake sequences were already assigned to long temporal correlations by measurements using conditional probabilities \cite{Bunde1,Livina1}.

Another example of naturally occurring bursty behavior is provided by the firing patterns of single neurons (see Materials and Methods). The recorded neural spike sequences display correlated and strongly inhomogeneous temporal bursty behavior, as shown in Fig.\ref{fig:2}.b. The distributions of the length of neural spike trains are found to be fat-tailed and indicate the presence of correlations between consecutive bursty spikes of the same neuron.

\subsubsection{Memory process}

In each studied system (communication of individuals, earthquakes at a given location, or single neurons) qualitatively similar behavior was detected as the single entities performed independent events or they passed through longer correlated bursty cascades. These cascades can be viewed as the result of building up some stress in the system (even at the single component level) e.g. the accumulation of important tasks in human communication, mechanical stress preceding an earthquake or integrated stimuli to a neuron. Our hypothesis is that each bursty period is related to a single excitation that triggers a cascade relieving this stress, which would naturally explain the correlation between the events within the same train. This ``one cascade-one task'' hypothesis can be relevant for human communication, where we speculate that a single task execution can occasionally trigger a series of events responsible for the appearance of long bursty trains. This assumption is supported by the observation that bursty nodes communicate predominantly with only one of their neighbors \cite{Kovanen2,Wu1}, indicating that a bursty period maybe linked to a single task. For earthquakes we can make the same conjecture, since a bursty period of earthquakes at a given time and location are most likely related to the same seismic activity. In case of neurons, the firings take place in bursty spike trains when the neuron receives excitatory input and its membrane potential exceeds a given potential threshold \cite{Nicholls1}. The spikes fired in a single train are correlated since they are the result of the same excitation and their firing frequency is coding the amplitude of the incoming stimuli \cite{Kandel1}.

\begin{figure*}[ht!] \centering
\includegraphics[width=160mm]{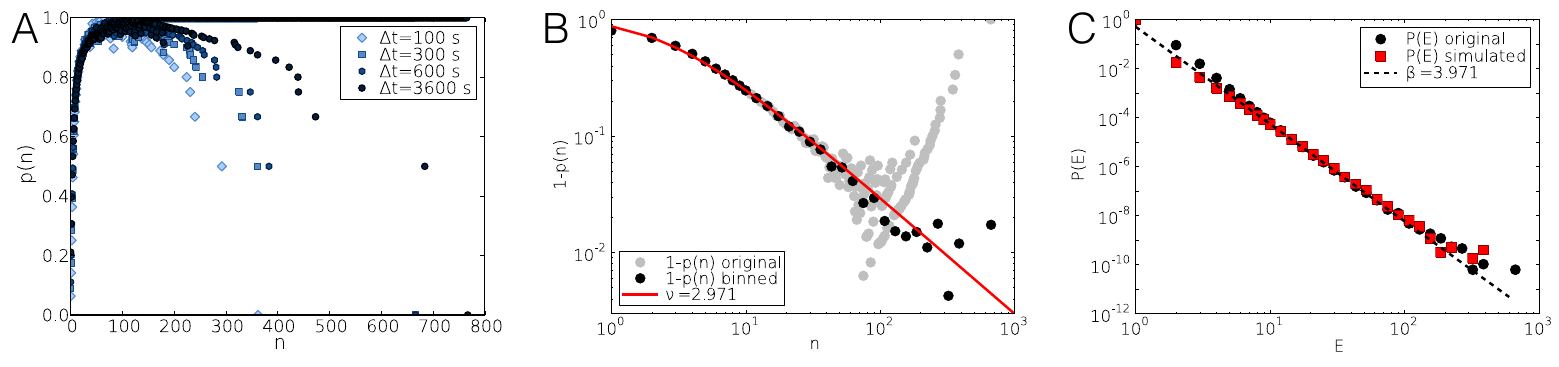}
\caption{Empirical and fitted memory functions of the mobile call sequence (a) Memory function calculated from the mobile call sequence using different $\Delta t$ time windows. (b) $1-p(n)$ complement of the memory function measured from the mobile call sequence with $\Delta t=600$ second and fitted with the analytical curve defined in equation (\ref{cmm:eq1}) with $\nu=2.971$. Grey symbols are the original points, while black symbols denotes the same function after logarithmic binning. (c) $P(E)$ distributions measured in real and in modeled event sequences.}
\label{figmm3:cond1}
\end{figure*}

The correlations taking place between consecutive bursty events can be interpreted as a memory process, allowing us to calculate the probability that the entity will perform one more event within a $\Delta t$ time frame after it executed $n$ events previously in the actual cascade. This probability can be written as:
\begin{equation}
 p(n)=\dfrac{\sum_{E=n+1}^{\infty}P(E)}{\sum_{E=n}^{\infty}P(E)}.
\label{EbS:eq1_1}
\end{equation}
Therefore the memory function, $p(n)$, gives a different representation of the distribution $P(E)$. The $p(n)$ calculated for the mobile call sequence are shown in Fig.\ref{figmm3:cond1}.a for trains detected with different window sizes. Note that in empirical sequences for trains with size smaller than the longest train, it is possible to have $p(n)=1$ since the corresponding probability would be $P(E=n)=0$. At the same time due to the finite size of the data sequence the length of the longest bursty train is limited such that $p(n)$ shows a finite cutoff.

We can use the memory function to simulate a sequence of correlated events. If the simulated sequence satisfies the scaling condition in (\ref{eq:E}) we can derive the corresponding memory function by substituting (\ref{eq:E}) into (\ref{EbS:eq1_1}), leading to:
\begin{equation}
 p(n) = \left( \frac{n}{n+1}\right) ^\nu
\label{cmm:eq1}
\end{equation}
with the scaling relation (see SI):
\begin{equation}
\beta=\nu +1.
\label{cmm:eq5}
\end{equation}

In order to check whether (\ref{cmm:eq5}) holds for real systems and whether the memory function in (\ref{cmm:eq1}) describes correctly the memory in real processes we compare it to a memory function extracted from an empirical $P(E)$ distributions. We selected the $P(E)$ distribution of the mobile call dataset with $\Delta t=600$ second and derived the corresponding $p(n)$ function. The complement of the memory function, $1-p(n)$, is presented in Fig.\ref{figmm3:cond1}.b where we show the original function with strong finite size effects (grey dots) and the same function after logarithmic binning (black dots).

Taking equation (\ref{cmm:eq1}) we fit the theoretical memory function to the log-binned empirical results using least-squares method with only one free parameter, $\nu$. We find that the best fit offers an excellent agreement with the empirical data (see Fig.\ref{figmm3:cond1}.b and also Fig.\ref{figmm3:cond1}.a) with $\nu=2.971\pm 0.072$. This would indicate $\beta\simeq 3.971$ through (\ref{cmm:eq5}), close to the approximated value $\beta\simeq 4.1$, obtained from directly fitting the empirical $P(E)$ distributions in the main panel of Fig.\ref{fig:1}.a (for fits of other datasets see SI). In order to validate whether our approximation is correct we take the theoretical memory function $p(n)$ of the form (\ref{cmm:eq1}) with parameter $\nu=2.971$ and generate bursty trains of $10^8$ events. As shown in Fig.\ref{figmm3:cond3}.c, the scaling of the $P(E)$ distribution obtained for the simulated event trains is similar to the empirical function, demonstrating the validity of the chosen analytical form for the memory function.

\subsection*{Model study}

As the systems we analysed are of quite different nature, from physics (earthquakes) to social (human communication) and biological (neuron spikes) systems, finding a single mechanistic model to fit them all is virtually impossible. Therefore, our goal is to define a phenomenological model that captures common features of the observed dynamics and see how these features are related to each other.

\subsubsection{Reinforcement dynamics with memory}

\begin{figure}[h!] \centering
\includegraphics[width=78mm]{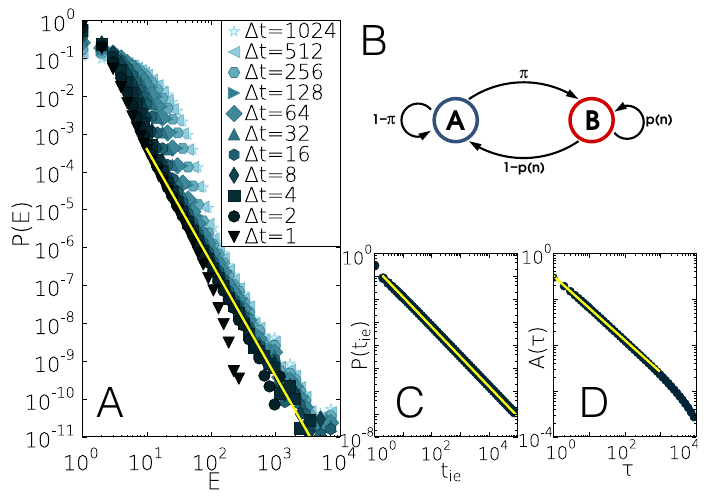}
\caption{Schematic definition and numerical results of the model study. (a) $P(E)$ distributions of the synthetic sequence after logarithmic binning with window sizes $\Delta t=1...1024$. The fitted power-law function has an exponent $\beta=3.0$. (b) Transition probabilities of the reinforcement model with memory. (c) Logarithmic binned inter-event time distribution of the simulated process with a maximum inter-event time $t_{ie}^{max}=10^6$. The corresponding exponent value is $\gamma=1.3$.  (d) The average logarithmic binned autocorrelation function with a maximum lag $\tau^{max}=10^4$. The function can be characterized by an exponent $\alpha=0.7$. Simulation results averaged over $1000$ independent realizations with parameters $\mu_A=0.3$, $\mu_B=5.0$, $\nu=2.0$, $\pi=0.1$ and $T=10^9$. For the calculation we chose the maximum inter-event time $t_{ie}^{max}=10^5$, which is large enough not to influence short temporal behavior, but it increases the program performance considerably.}
\label{figmm3:cond3}
\end{figure}

We assume that the investigated systems can be described with a two-state model, where an entity can be in a normal state $A$, executing independent events with longer inter-event times, or in an excited state $B$, performing correlated events with higher frequency, corresponding to the observed bursts. To induce the inter-event times between the consecutive events we apply a reinforcement process based on the assumption that the longer the system waits after an event, the larger the probability that it will keep waiting. Such dynamics shows strongly heterogeneous temporal features as discussed in \cite{Stehle:2010,Zhao:2011}. For our two-state model system we define a process, where the generation of the actual inter-event time depends on the current state of the system. The inter-event times are induced by the reinforcement functions that give the probability to wait one time unit longer after the system has waited already time $t_{ie}$ since the last event. These functions are defined as
\begin{equation}
 f_{A,B}(t_{ie})=\left( \dfrac{t_{ie}}{t_{ie}+1} \right)^{\mu_{A,B}}
 \label{eq:rfunc1}
\end{equation}
where $\mu_A$ and $\mu_B$ control the reinforcement dynamics in state $A$ and $B$, respectively. These functions follow the same form as the previously defined memory function in (\ref{cmm:eq1}) and satisfy the corresponding scaling relation in (\ref{cmm:eq5}). If $\mu_A\ll \mu_B$ the characteristic inter-event times at state $A$ and $B$ become fairly different, which induces further temporal inhomogeneities in the dynamics. The actual state of the system is determined by transition probabilities shown in Fig.\ref{figmm3:cond3}.b, where to introduce correlations between consecutive excited events performed in state $B$ we utilize the memory function defined in equation (\ref{cmm:eq1}).

To be specific, the model is defined as follows: first the system performs an event in a randomly chosen initial state. If the last event was in the normal state $A$, it waits for a time induced by $f_A(t_{ie})$, after which it switches to excited state $B$ with probability $\pi$ and performs an event in the excited state, or with probability $1-\pi$ stays in the normal state $A$ and executes a new normal event. In the excited state the inter-event time for the actual event comes from $f_B(t_{ie})$ after which the system decides to execute one more excited event in state $B$ with a probability $p(n)$ that depends on the number $n$ of excited events since the last event in normal state. Otherwise it switches back to a normal state with probability $1-p(n)$. Note that a similar model without memory was already defined in the literature \cite{Kleinberg1}. 

The numerical results predicted by the model are summarized in Fig.\ref{figmm3:cond3} and Table \ref{table:1}. We find that the inter-event time distribution in Fig.\ref{figmm3:cond3}.c reflects strong inhomogeneities as it takes the form of a scale-free function with an exponent value $\gamma=1.3$, satisfying the relation $\gamma=\mu_A+1$. As a result of the heterogeneous temporal behavior with memory involved, we detected spontaneously evolving long temporal correlations as the autocorrelation function shows a power-law decay. Its exponent $\alpha=0.7$ (see Fig.\ref{figmm3:cond3}.d) also satisfies the relation $\alpha+\gamma=2$ (see SI). The $P(E)$ distribution also shows fat-tailed behavior for each investigated window size ranging from $\Delta t=1$ to $2^{10}$ (see Fig.\ref{figmm3:cond3}.a). The overall signal here is an aggregation of correlated long bursty trains and uncorrelated single events. This explains the weak $\Delta t$ dependence of $P(E)$ for larger window sizes, where more independent events are merged with events of correlated bursty cascades, which induces deviation of $P(E)$ from the expected scale-free behavior. The $P(E)$ distributions can be characterized by an exponent $\beta=3.0$ in agreement with the analytical result in (\ref{cmm:eq5}) and it confirms the presence of correlated bursty cascades. In addition, even if we fix the value of $\beta$ and $\gamma$, the $\alpha$ exponent satisfies the condition $\alpha < \gamma < \beta$, an inequality observed in empirical data (see Table \ref{table:1}).

\section{Discussion}

In the present study we introduced a new measure, the number of correlated events in bursty cascades, which detects correlations and heterogeneity in temporal sequences. It offers a better characterization of correlated heterogeneous signals, capturing a behavior that cannot be observed from the inter-event time distribution and the autocorrelation function. The discussed strongly heterogeneous dynamics was documented in a wide range of systems, from human dynamics to natural phenomena. The time evolution of these systems were found to be driven by temporal correlations that induced scale-invariant distributions of the burst lengths. This scale-free feature holds for each studied systems and can be characterized by different system-dependent exponents, indicating a new universal property of correlated temporal patterns emerging in complex systems.

We found that the bursty trains can be explained in terms of memory effects, which can account for the heterogeneous temporal behavior. In order to better understand the dynamics of temporally correlated bursty processes at single entity level we introduced a phenomenological model that captures the common features of the investigated empirical systems and helps us understand the role they play during the temporal evolution of heterogeneous processes.

\section{Materials and Methods}

\textbf{Data processing.} To study correlated human behavior we selected three datasets containing time-stamped records of communication through different channels for a large number of individuals. For each user we extract the sequence of outgoing events as we are interested in the correlated behavior of single entities. The datasets we have used are as follows: (a) A mobile-call dataset from a European operator covering $\sim 325\times 10^6$ million voice call records of $\sim 6.5\times 10^6$ users during $120$ days \cite{Karsai1}. (b) Text message records from the same dataset consisting of $125.5\times 10^6$ events between the same number of users. Note that to consider only trusted social relations these events were executed between users who mutually called each other at least one time during the examined period. Consecutive text messages of the same user with waiting times smaller than $10$ seconds were considered as a single multipart message \cite{Kovanen1} though the $P(t_{ie})$ and $A(\tau)$ functions do not take values smaller than $10$ seconds in Fig.\ref{fig:1}.b. (c) Email communication sequences of $2,997$ individuals including $20.2\times 10^4$ events during $83$ days \cite{Eckmann1}. From the email sequence the multicast emails (consecutive emails sent by the same user to many other with inter-event time 0) were removed in order to study temporally separated communication events of individuals. To study earthquake sequences we used a catalog that includes all earthquake events in Japan with magnitude larger than two between 1986 and 1998 \cite{JEQ1}. We considered each recorded earthquake as a unique event regardless whether it was a main-shock or an after-shock. For the single station measurement we collected a time order list of earthquakes with epicenters detected at the same region \cite{Bak1,Zhao1} (for other event collection methods see SI). The resulting data consists of $198,914$ events at $238$ different regions. The utilized neuron firing sequences consist of $31,934$ outgoing firing events of $1,052$ single neurons which were collected with $2$ millisecond resolution from rat's hippocampal slices using fMCI techniques \cite{Ikegaya1,Takahashi1}.

\textbf{Inter-event time shuffling of real sequences.} For the independent event sequences in Fig.\ref{fig:1} and \ref{fig:2} (empty symbols) we shuffled the inter-event times of individuals allowing to change the inter-event time values between any users but keeping the original event number for each individual. The presented $P(E)$ distributions were calculated with one $\Delta t$ window size to demonstrate the exponential behavior of $P(E)$ for independent events.

\begin{acknowledgments}
We thank J. Saram\"aki, H-H. Jo, M. Kivel\"a, C. Song and D. Wang for comments and useful discussions. Financial support from EU’s FP7 FET-Open to ICTeCollective Project No. 238597 and TEKES (FiDiPro) are acknowledged.
\end{acknowledgments}

\onecolumngrid
\newpage

\begin{center}
\LARGE
Universal features of correlated bursty behaviour \\ Supplementary Informations \\
\large
\vspace{.3in}
M. Karsai, K. Kaski, A.-L. Barab\'asi and J. Kert\'esz
\vspace{.3in}
\end{center}

\normalsize
 
\maketitle

\section{Scaling of the autocorrelation function in heterogeneous independent processes}

\begin{figure}[ht!] \centering
\includegraphics[width=160mm]{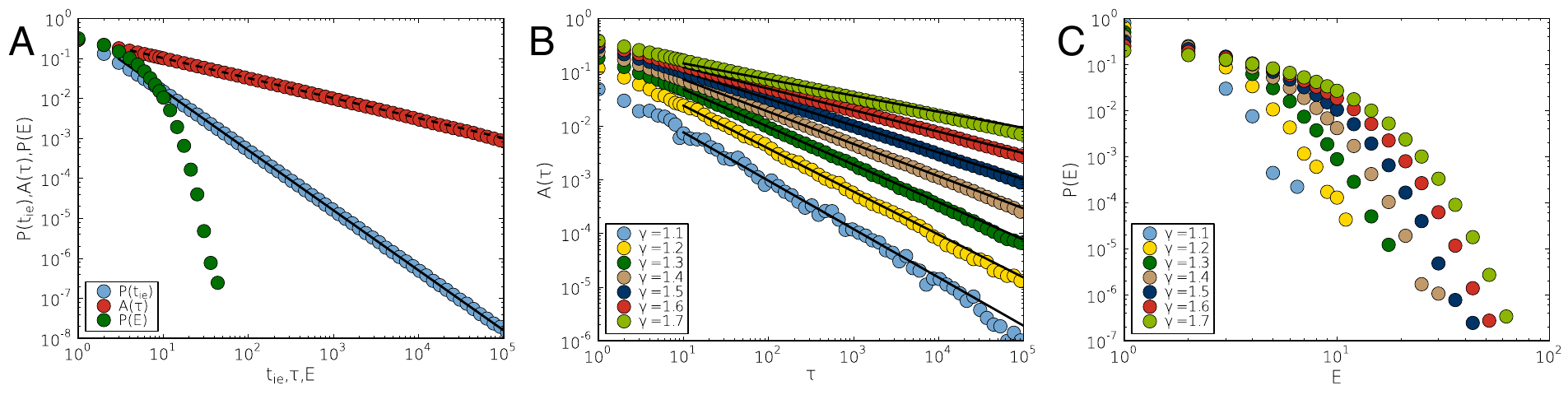}
\caption{\textbf{The characteristic functions calculated for heterogeneous independent signals.} (a) $P(t_{ie})$, $A(\tau)$ and $P(E)$ functions for $\gamma=1.5$. Solid line is a power-law function with the given $\gamma$ exponent value, while dashed line denotes a a power-law function with an effective $\alpha=0.5$ exponent value. (b) $A(\tau)$ effective autocorrelation functions for various $\gamma$ exponents. Straight lines are denoting power-law functions with $\alpha$ exponents satisfying the $\alpha+\gamma=2$ relation. (c) Corresponding $P(E)$ distributions for various $\gamma$ exponents.}
\label{figIM:3}
\end{figure}

As it was mentioned in the main text for strongly inhomogeneous temporal sequences of independent events the Hurst exponent can take a value $H>1/2$ and can assign false positive temporal correlations \cite{Hansen1}. This effective behaviour is also reflected by the autocorrelation function which can show power-law scaling $A(\tau)\sim \tau^{-\alpha}$ for sequences of independent events with inter-event time distribution $P(t_{ie})\sim t_{ie}^{-\gamma}$, and can indicate presence of non-existing correlations. Based on the generating function method and the Tauberian theorems one can show that for $1\leq \gamma \leq 2$ the scaling law 
\begin{equation}
\alpha + \gamma = 2
\label{Leq:3}
\end{equation}
holds \cite{Vajna1}. We checked this relationship by numerical simulations, where we generate independent events with inter-event times sampled from a power-law distribution with a fixed exponent $\gamma$ and calculate the autocorrelation function. As it is shown in Fig.\ref{figIM:3}.a the exponent relation in Eq.\ref{Leq:3} holds for numerical results since for $P(t_{ie})$ with $\gamma=1.5$ (blue symbols and straight line) the effective autocorrelation function (red symbols) scales as a power-law with an exponent $\alpha=0.5$ (dashed line). At the same time the $P(E)$ distribution calculated with $\Delta t=10$ window size (green symbols) indicates the true uncorrelated behaviour as it shows exponential decay.

The same calculations had been repeated for various $1\leq \gamma \leq 2$ values averaged over 1000 independent realization as we present it in Fig.\ref{figIM:3}.b. In each cases the autocorrelation function satisfies the condition derived in Eq.\ref{figDS:1}. However, as $\gamma \rightarrow 1$ extreme fluctuations start to influence the dynamics considerably, while some discrepancy appears as $\gamma \rightarrow 2$ where fully correlated behaviour should evolve which cannot be the case for random processes even the fluctuations are finite.	

The corresponding $P(E)$ distributions show exponential behaviour as it is demonstrated in Fig.\ref{figIM:3}.c and assign the true uncorrelated temporal behaviour of the processes. It demonstrates that autocorrelation is unable to address present correlations obviously for heterogeneous temporal processes since it indicates effective correlations between independent events. However, the $P(E)$ distribution is capable to detect correlated behaviour even for processes with fat-tailed inter-event time distributions as it decays exponentially in case of independent signals (for a detailed study see SI Section \ref{sec:indMod}) while it scales as a power-law when long bursty periods evolve as a result of temporal correlations.

\section{P(E) distributon in independent models}
\label{sec:indMod}

We study the functional behavior of the $P(E)$ bursty event number distribution for processes of independent events. As it was already discussed in the main text the $P(E)$ distribution with a given $\Delta t$ time window size can be written as:
\begin{equation}
P(E=n)= \left(  \int_0^{\Delta t}P(t_{ie})dt_{ie} \right)^{n-1} \left( 1-\int_0^{\Delta t}P(t_{ie})dt_{ie} \right)
\end{equation}
for $n>0$ in case of processes where we sample $t_{ie}$ inter-event times independently from an arbitrary $P(t_{ie})$ distribution. If the integral $\int_0^{\Delta t}P(t_{ie})dt_{ie}<1$, then $P(E=n)\sim a^{-(n-1)}$ is decreasing exponentially, otherwise if $\int_0^{\Delta t}P(t_{ie})dt_{ie}=1$ then $P(E=n)=0$. Since we fix the upper limit of the integrand $\Delta t < \infty$, the first behavior holds.

To numerically confirm this analytical results we define a single user model where we generate events with inter-event times sampled from two different distributions:
\begin{equation}
 P(t_{ie})\sim t^{-\gamma} \mbox{\hspace{.3in} and \hspace{.3in}} P(t_{ie})\sim e^{-t/\tau}
\end{equation}
and count the number of consecutive events which fall into a bursty periods with $t_{ie}\leq \Delta t$.

\begin{figure}[ht!] \centering
\includegraphics[width=120mm]{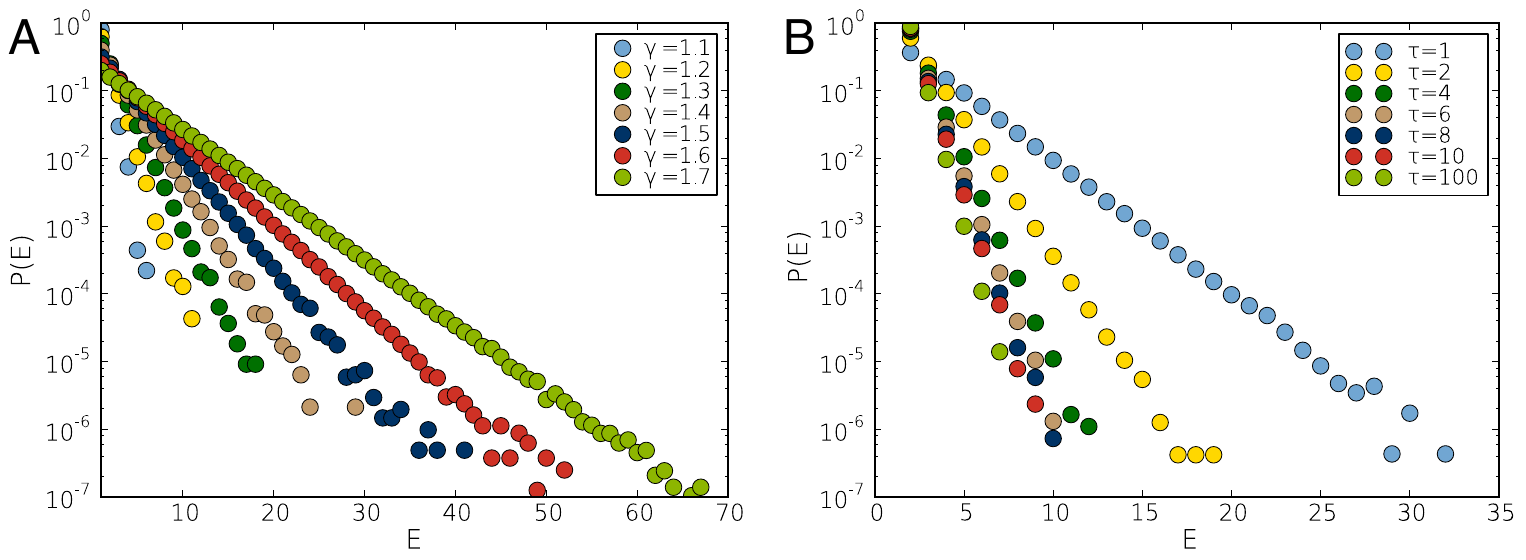}
\caption{\textbf{P(E) distributor in independent models.} Distribution of number of bursty events in periods evolved in independent processes with (a) power-law and (b) exponential inter-event  time distributions. Simulation results are presented with various parameter values and with fixed $\Delta t$ and maximum inter-event times.}
\label{figIM:1}
\end{figure}

The simulation results in Fig.\ref{figIM:1} demonstrates the predicted exponential decay for the $P(E)$ distributions. We performed sequences with various $\gamma$ or $\tau$ values, but the $P(E)$ distribution was calculated in each cases with a fixed $\Delta t$ window size.

\section{$\Delta t$ time window dependence of $P(E)$ distribution}

\begin{figure*}[h!] \centering
\includegraphics[width=160mm]{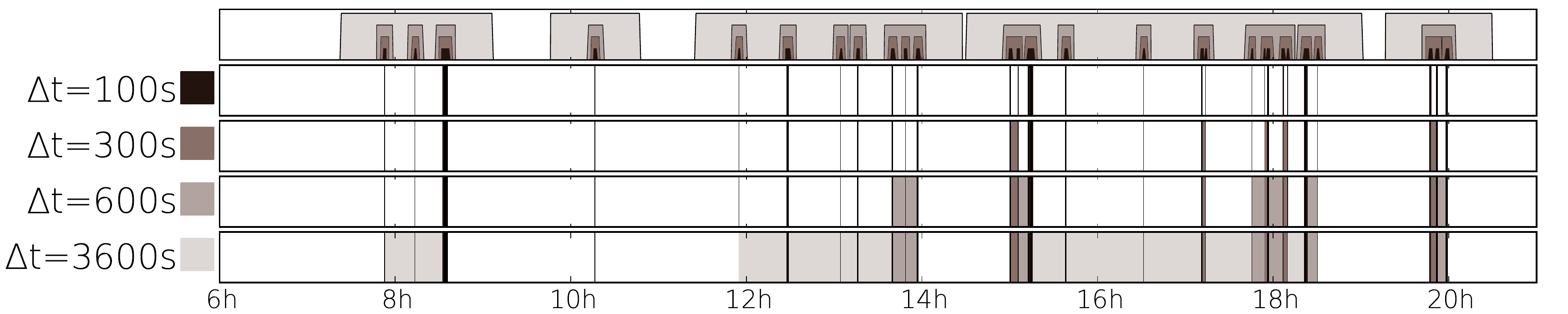}\\
\caption{\textbf{Illustration of bursty period detection in the call sequence of a selected individual.} Black spikes denote the call events with width rational to the call length. In the 2-5 lines we colored the inter-event periods if they were smaller then the corresponding time window size. As we are increasing $\Delta t$ the periods which were detected with smaller time-windows are be merged together to longer trains. The first line demonstrates the fine grained self-containing structure of the long evolving bursty periods.}
\label{fig:schem3}
\end{figure*}

\begin{figure}[ht!] \centering
  \includegraphics[width=92mm]{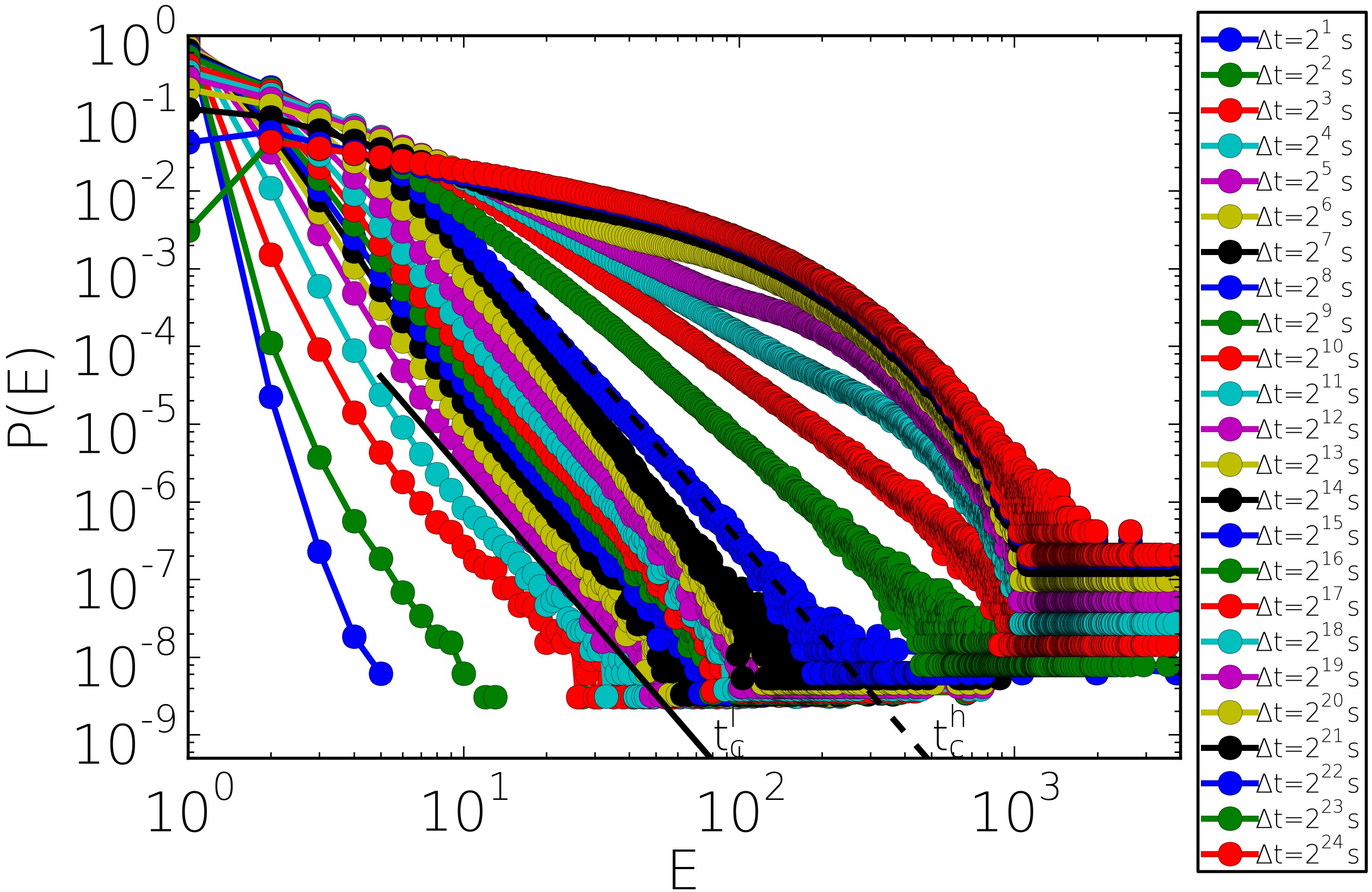}
\caption{\textbf{Distribution of correlated bursty event numbers with various time-window size.} $P(E)$ distribution calculated for various $\Delta t$ window size with value varying between $\Delta t=2^1...2^{24}$. Straight lines are denoting the lower characteristic time $t_c^l=20$ seconds (solid line) and the higher characteristic time $t_c^h=12$ hours (dashed line) and assign power-law functions with exponent $\beta\simeq 4.1$.}
\label{figSD:3}
\end{figure}

We have seen in the main text that the $P(E)$ bursty train size distribution shows a robust behaviour against the selected $\Delta t$ time windows. By increasing $\Delta t$ the correlated event clusters are growing as more and more single events and periods with shorter time windows get merged together as depicted in Fig.\ref{fig:schem3}. Looking for a wider range of $\Delta t$ from $1$ second (the time resolution of the mobile-call sequences) up to $T=120$ days (the length of the data sequence) we can follow how $P(E)$ distribution evolves. In Fig.\ref{figSD:3}.a we present $P(E)$ distributions with $\Delta t=2^n$ where $n$ goes from $1$ to $24$. A lower $t_c^l$ and a higher $t_c^h$ characteristic time can be detected by looking at the evolution of $P(E)$:
\begin{itemize}
\item if $\Delta t < t_c^l$: only small bursty trains evolve and the $P(E)$ distributions show a concave behaviour.
\item if $t_c^l < \Delta t < t_c^h$: critical regime where the $P(E)$ distributions are scaling as a power-law with an exponent $\beta$ insensitive for $\Delta t$.
\item if $\Delta t > t_c^h$: Uncorrelated periods are merged together and $P(E)$ is approaching the strength distribution as $\Delta t\rightarrow T$ days where for each user only one bursty period evolves containing all events of the present user.
\end{itemize}
For the mobile call sequence the corresponding exponent takes $\beta=4.1$ and the two characteristic times are $t_c^l=20$ seconds and $t_c^h=12$ hours. These times are also play crucial roles for $P(t_{ie})$ and $A(\tau)$ as $20$ seconds is the typical reaction time between two consecutive call actions, while $12$ hours reflects the length of correlated periods of human daily activity. Looking at the scaling behaviour of $P(t_{ie})$ and $A(\tau)$ of other human activities, earthquakes or neuron spike sequences, one can detect analogous evolving behaviour of $P(E)$ with corresponding characteristic times.

To exclude the possibility that the broad strength distribution causes the power law like $P(E)$ behaviour, we repeated the same measurement with a single event sequence which was constructed from the call sequences of $10^6$ users and monitor how $P(E)$ changes as we increase $\Delta t$ from its minimum to its maximum (not shown here). In this case we also observed a extended critical regime where $P(E)$ presented scale-free behaviour with the same exponent $\beta\simeq 4.1$.

\section{Strength decomposition of the characteristic functions}

We demonstrated in the main text that bursty correlated behavior can be characterized by functions as $P(t_{ie})$, $A(\tau)$ and $P(E)$ in several kind of dynamic systems. However, the question remained whether the observed scaling behaviour of these functions is an artifact of other present inhomogeneities e.g. the broad distribution of node activity (strength), or it de facto reflects general behavioural characteristics of entities. To answer this question we completed additional measurements on the mobile-call dataset which possesses large enough size to provide good statistics after we decompose users into different groups. We also repeated the following measurements for the other datasets and found similar behaviour of the analyzed functions.

\begin{figure}[ht!] \centering
\includegraphics[width=120mm]{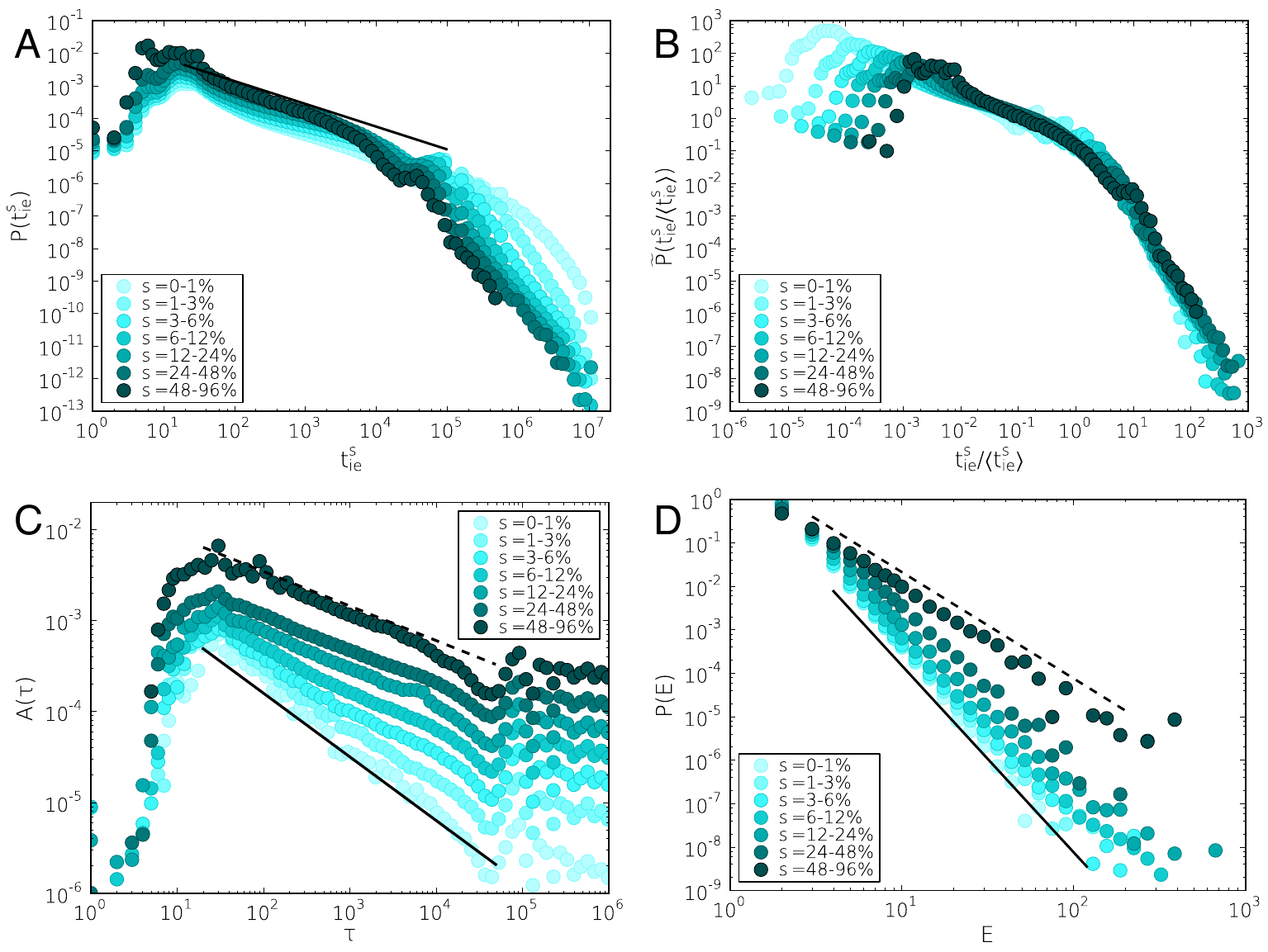}
\caption{\textbf{Characteristic functions of different strength groups.} (a) Inter-event time distributions measured between the outgoing calls of individuals belonging to the same strength group (see text). The solid line assigns a power-law function with exponent $\gamma=0.7$. (b) The same curves scaled by using the $\langle t^s_{ie} \rangle P(t^s_{ie})= \widetilde{P}(t^s_{ie}/\langle t^s_{ie} \rangle)$ where $t^s_{ie}$ denotes the inter-event times of individuals with a given $s$ strength and $\langle t^s_{ie} \rangle$ assigns the average inter-event time calculated for users with strength $s$. (c) Autocorrelation function averaged for users belonging to different strength groups. Characteristic exponents are changing between $\alpha=0.7$ (solid line) and $\alpha=0.38$ (dashed line) as the strength of the corresponding groups is increasing. (d) $P(E)$ distribution for users belonging to different strength groups. The corresponding exponents are changing between $\beta=2.45$ (dashed line) and $\gamma=4.3$ (solid line).}
\label{figSD:2}
\end{figure}

Here we define node strength $s$ as the number of (in and out) calls made by a user during the entire period. It is known from earlier studies \cite{Onnela1} that the $P(s)$ strength distribution of the MCD is broad with a maximum value $s_{max}=7933$ in the present case. Take into consideration the inhomogeneous $P(s)$ distribution we ranked users into strength groups with increasing bin size. We realized from individual level analysis that users with the largest strength values are playing a disparate role as they show non-human like communication patterns. Though we exclude them from the measurements and take users only with strength smaller then $96\%$ of the maximum value.

In Fig.\ref{figSD:2} we present the average characteristic functions calculated separately for each strength groups. Heterogeneous inter-event time distributions are characterizing the communication of users ranked into different groups (Fig.\ref{figSD:2}.a). Only the tail of the $P(t_{ie})$ distributions show significant discrepancy due to different activity levels. Since users frequently show correlated bursts with short inter-event times, those ones with small number of calls in sum have longer inactive periods between bursts which induces a longer tail in $P(t_{ie})$ with a later exponential cutoff. Users with higher activity have shorter waiting times between bursts which is reflected by the earlier turning point of the inter-event time distribution. In order to proof whether these distributions are broad not only due to the inhomogeneous strength distributions in Fig.\ref{figSD:2}.b we scaled them together using the scaling relation $\langle t^s_{ie} \rangle P(t^s_{ie})= \widetilde{P}(t^s_{ie}/\langle t^s_{ie} \rangle)$ where $\langle t^s_{ie} \rangle$ denoted the average inter-event times, calculated separately for each groups as it was done in \cite{Karsai1,Candia1}. Using this scaling relation the distributions scale together on the same master curve which indicates that $P(t_{ie})$ follows the same functional behaviour independently from the chosen strength group (and the average inter-event time of this group).

The autocorrelation function in Fig.\ref{figSD:2}.c decays as a power-law up to $12$ hours assigning long-temporal correlations for users in each strength groups. Naturally stronger correlations are detected for more active users as the corresponding exponents vary between $\alpha\simeq 0.38...0.7$ as we decrease the activity level. A similar behaviour is confirmed by the scaling of $P(E)$ in Fig.\ref{figSD:2}.d as the distributions (calculated with $\Delta t=600$ seconds) remain fat-tailed for each user groups with an exponent between $\beta=2.45...4.3$ as we decrease activity. It implies that long correlated bursty periods evolve even for users with only a few calls but with smaller probability then for users with many call actions.

\section{De-seasoned results}

In order to study the effect of circadian patterns on the observed behaviour of the characteristic functions in mobile call communication, we remove daily fluctuations by rescaling the event times using a method defined in \cite{Jo1,Anteneodo1}. Measuring the $r(t)$ event density and its average value $R_t$ during the entire period we can define a rescaled time for each event as:
\begin{equation}
d t^{*}= \frac{r(t)}{R_t}dt=\rho(t)dt
\end{equation}
where $\rho(t)$ denotes the event rate and the $\rho^*(t^*)dt^*=\rho(t)dt$ scaling holds for the time variable with $\rho^*(t^*)=1$. Rescaling event times with these conditions, events in periods with high event frequency become dilated, while event times are contracted when their frequency is lower, so the effect of cyclic fluctuations can be reduced. We rescaled the times of the outgoing call events of each user in the MCD considering the duration of their calls and ranked them into strength groups using their overall call activity.

\begin{figure}[ht!] \centering
\includegraphics[width=120mm]{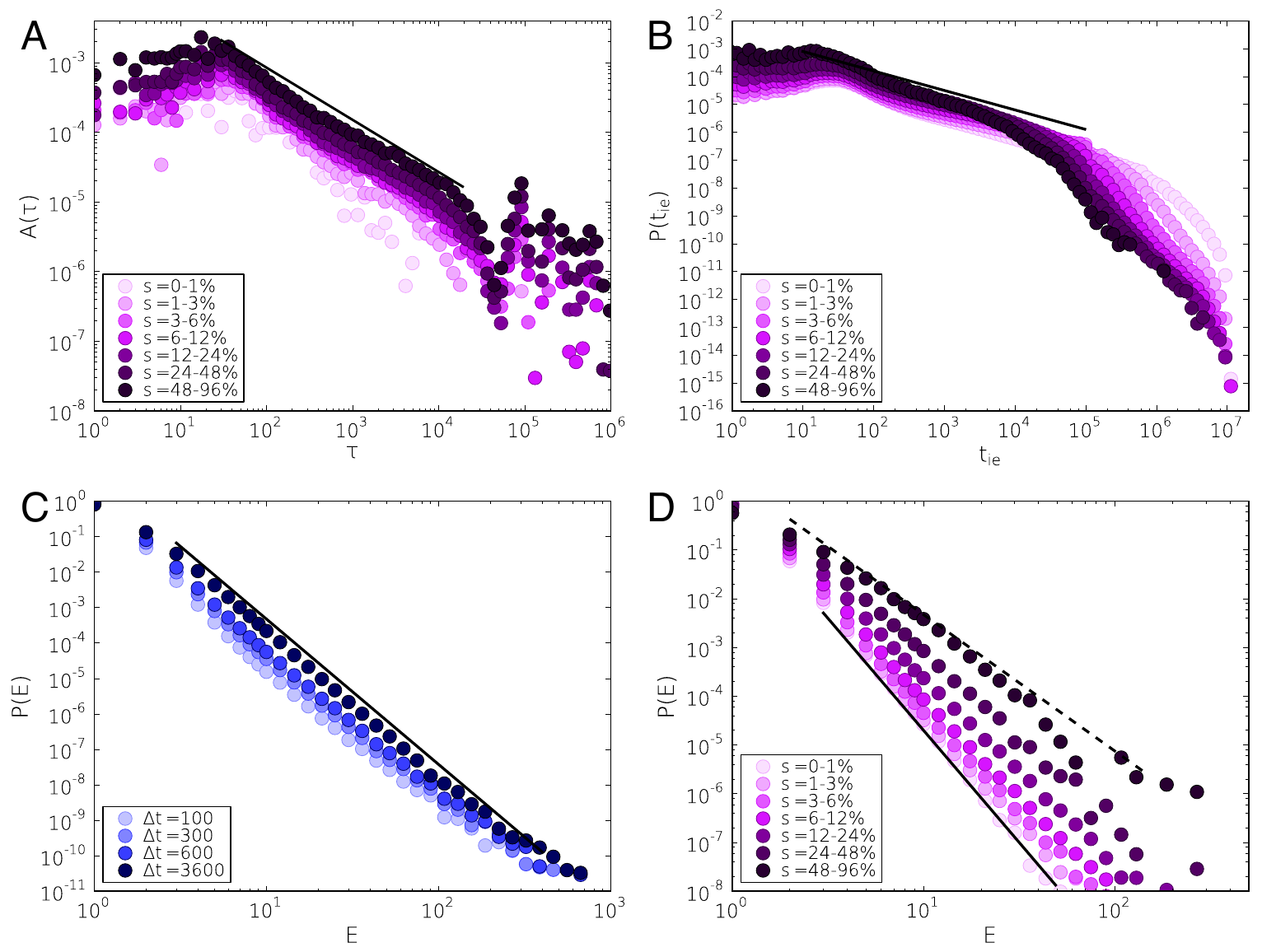}
\caption{\textbf{Characteristic functions of de-seasoned sequences.} (a) Autocorrelation functions and (b) inter-event time distributions of de-seasoned outgoing call event sequences of individuals belonging to the same strength groups. Dashed line on panel (a) assigns a power-law function with exponent $\alpha= 0.65$ while on panel (b) with exponent $\gamma=0.7$. (c) $P(E)$ distributions measured with various $\Delta t$ time window size in the de-seasoned outgoing event sequences of individuals. The slope of the dashed straight line indicates an exponent value $\beta=4.1$. (d) $P(E)$ distribution with $\Delta t=600$ measured for users belonging to different strength groups. The solid (dashed) line denotes a power-law function with exponent $4.6$ ($2.8$.)}
\label{figDS:1}
\end{figure}

Utilizing the sequence of outgoing call events with rescaled times we calculate the three characteristic functions to see the impact of daily fluctuations on them. Since cyclic fluctuations are introducing correlations in the dynamics, by removing them the $A(\tau)$ function should reflect weaker correlations and if only the circadian patterns are responsible for the temporal correlations this function should radically change and present short term correlated behaviour only. In Fig.\ref{figDS:1}.a the autocorrelation denotes reduced correlated behaviour compare to the same function of the original event sequence in Fig.\ref{figSD:2}.c, however it signifies remaining long temporal correlations as it shows slow decay with a slightly larger exponent $\alpha\simeq 0.75$ compared to the unscaled value ($\alpha=0.5$). The inter-event time distributions in Fig.\ref{figDS:1}.b also remains similar compared to Fig.\ref{figSD:2}.a with approximately the same exponent value $\gamma\simeq 0.7$. It is in complete agreement with the results presented in \cite{Jo1} where the $P(t_{ie})$ of call events remained unchanged after the same de-seasoning method was applied on the event sequence.

Long bursty periods are not destroyed by removing daily patterns from the event sequence as it is shown in Fig.\ref{figDS:1}.c. The long periods of outgoing bursts of individuals remain for various $\Delta t$ values reflected by power-law like $P(E)$ distributions decreasing with exponent $\beta\simeq 4.1$ similar to the original sequence. The $P(E)$ distribution becomes more disperse if we decompose it by strength for periods with time window size $\Delta t=600$ seconds (see Fig.\ref{figDS:1}.d). The tail of the decomposed $P(E)$ distributions can be estimated with exponents between $\beta\simeq 2.8 ... 4.6$. Consequently as the scaling behaviour of the de-seasoned data show similar behaviour as the original sequence, it implies that even circadian fluctuations are partially responsible for the present correlations they do not effect considerably the evolution of long correlated bursty sequence in communication of individuals.

\section{Bursty-topology correlations in earthquake-sequences}

\begin{figure}[ht!] \centering
\includegraphics[width=160mm]{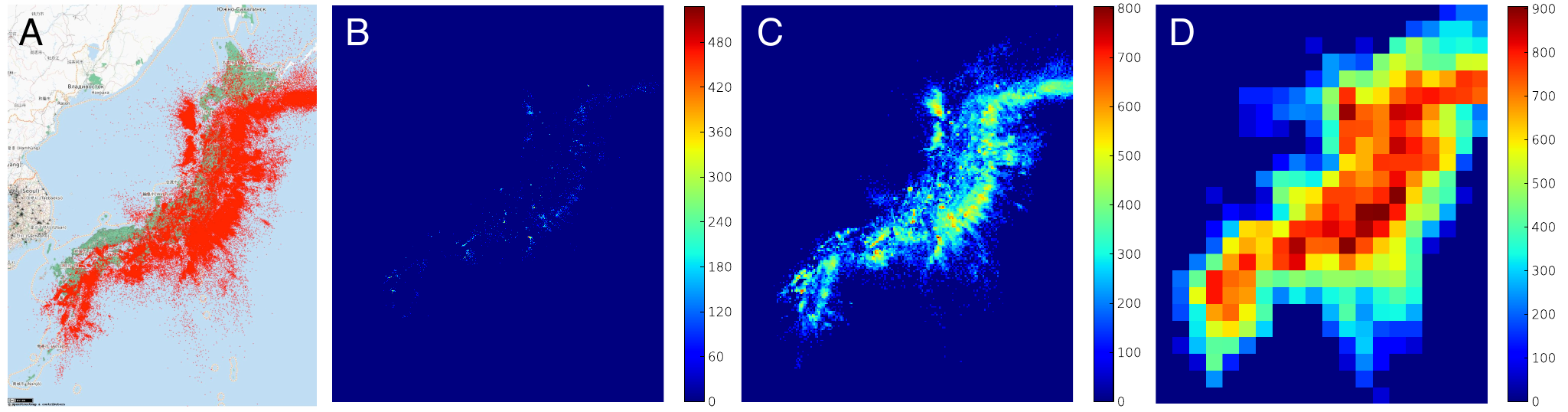}
\caption{(a) Earthquake events in Japan between 1985.07-1998.12. Count of earthquake events located into area bins with size (b) $10^0 km^2$, (c) $10^4 km^2$ and (d) $10^4 km^2$.}
\label{EQ:fig1}
\end{figure}

In order to measure temporal correlations between earthquake events we applied a so-called single-station method \cite{Smalley1,Udias1,Zhao1} and studied event sequences executed at the same geographical area. For each earthquake event the longitude and latitude coordinates of the epicenter was known from informations measured at several surrounding seismological stations \cite{JEQ1}. The distribution of epicenters on the investigated area is visualized on Fig.\ref{EQ:fig1}.a. 

For the results presented in the main text we grouped $198,914$ events into $238$ administrative regions and we observed scale-free behaviour for the characteristic functions (see Fig.3.a in the main text). However, since the regions can have different sizes the question remains whether the observed scaling behaviour is the result of the diverse size distribution of different regions or it truly assigns heterogeneous correlated temporal behaviour. To answer this question we divided the investigated area for bins with equal sizes and for each bin we collected a time-ordered list of events executed on the corresponding area. We used three different bin sizes: $10^0 km^2$, $10^2 km^2$ and $10^4 km^2$. The event count map with the three different bin sizes is shown in Fig.\ref{EQ:fig1}.b, c and d.

\begin{figure}[ht!] \centering
\includegraphics[width=160mm]{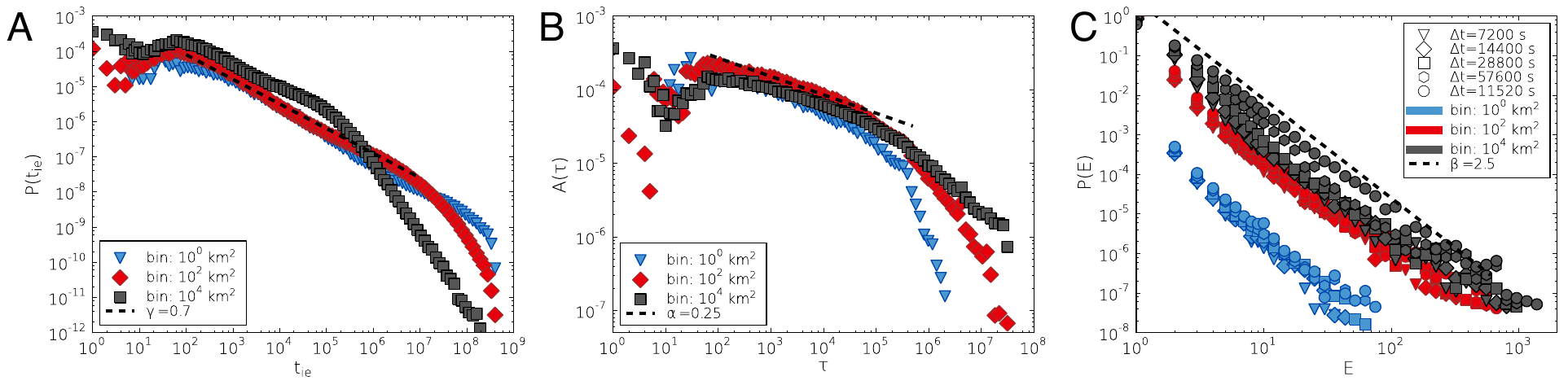}
\caption{Characteristic functions: (a) $P(t_{ie})$, (b) $A(\tau)$ and (c) $P(E)$ calculated for sequences collected with area bin sizes $10^0 km^2$ (blue), $10^2 km^2$ (red) and $10^4 km^2$ (grey). The autocorrelation functions on (b) were averaged for the first $100$ most active bins with the corresponding size. On (c) the different symbols denote the $P(E)$ distributions calculated with different $\Delta t$ time window sizes.}
\label{EQ:fig3}
\end{figure}

Using the event sequences collected for each bins we calculated the characteristic function $P(t_{ie})$, $A(\tau)$ and $P(E)$. As it is shown in Fig.\ref{EQ:fig3} each function remained fat tailed with the same exponent values $\beta=2.5$ and $\gamma=0.7$ as it was found for the measurements presented in the main text. The autocorrelation function exponent took a value $\alpha=0.25$ which assigns slightly stronger correlations as for the earlier calculations, however the average autocorrelation functions here were calculated for the first $100$ most active bins which explains the slight decrease of the $\alpha$ exponent.

As the scaling behaviour of the characteristic functions were re-found for calculations with equal-size area bins, in conclusion we can say that the observed temporal correlations and heterogeneous dynamics are not the result of the event collection method but they truly characterize the sequences of earthquake events.

\section{Calculation of the memory function}

In the main text for correlated event sequences with $P(E)\sim E^{-\beta}$ we can derive the memory function in the form:
\begin{equation}
p(n)=\left( \frac{n}{n+1} \right)^{\nu}
\label{eq:mf1}
\end{equation}
However, by only knowing Eq.\ref{eq:mf1} we can also calculate the related $P(E)$ distribution as:
\begin{eqnarray}
 P(E=n) &=& \left(1-\left(\frac{n}{n+1}\right)^\nu\right) \prod_{i=1}^{i=n-1}\left(\frac{i}{i+1}\right)^\nu\label{eq:mf2_1} = \left(1-\left(\frac{n}{n+1}\right)^{\nu}\right) \left(\frac{1}{n}\right)^{\nu} =\\
         &=& \frac{(n+1)^\nu-n^\nu}{(n+1)^\nu n^\nu} = \frac{(1+n^{-1})^\nu -1}{(n+1)^\nu} = \frac{\left( 1+\frac{\nu}{n}+\frac{\nu (\nu +1)}{2} \frac{1}{n^2}+... \right) -1}{(n+1)^\nu} \sim\\
         &\sim& \frac{\nu}{n(n+1)^{\nu}} \sim \frac{\nu}{n^{\nu +1}} \label{eq:mf2_2}
\end{eqnarray}
Consequently our calculations are consistent as the corresponding $P(E)$ distribution shows asymptotically a power-law decay with an exponent satisfying the relation:
\begin{equation}
\beta=\nu+1.
\label{eq:mf3}
\end{equation}

\section{Estimating exponent $\nu$ for empirical memory functions}

\begin{figure}[ht!] \centering
\includegraphics[width=160mm]{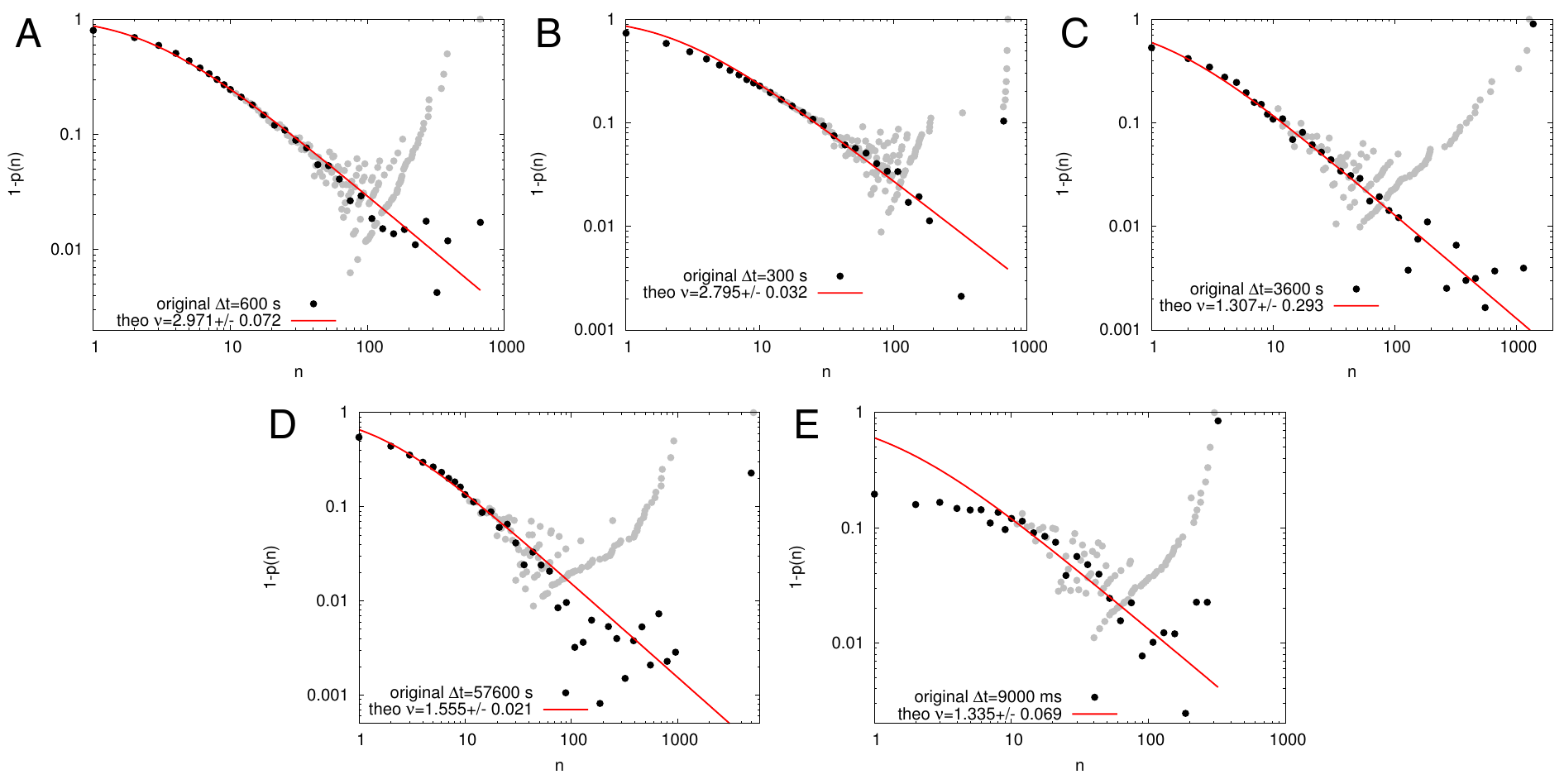}
\caption{\textbf{Fitted complement memory functions of different empirical data sequences calculated with a selected time-window size.} (a) Mobile call sequence ($\Delta t=600$ seconds). (b) Text message sequence ($\Delta t=300$ seconds). (c) Email sequence ($\Delta t=3600$ seconds). (d) Japanese earthquake sequence ($\Delta t=16$ hours). (e) Neuron firing sequence ($\Delta t=900$ milliseconds).}
\label{figIM:4}
\end{figure}

In order to estimate the $\nu$ memory function exponents for empirical sequences we calculated the $p(n)$ function for each investigated data sequence with a chosen $\Delta t$ time window size and plot the complement $1-p(n)$ of the original memory functions in Fig.\ref{figIM:4} (grey symbols) and the same functions after logarithmic binning (black symbols). As it is visible the original complement memory functions show strong finite size effects as the investigated event sequences span on a limited time frame and the related $P(E)$ distributions are finite. Similarly as it was discussed in the main text, we fitted the binned empirical memory functions with the analytical functions of the form of Eq.\ref{eq:mf1} using least-squares method with only one free parameter, the exponent $\nu$ (red lines in Fig.\ref{figIM:4}). The resulted $\nu$ exponents are written in the figure caption and also summarized in the main text in Table I. The derived exponents approximately satisfy the relation in Eq.\ref{eq:mf3} with the corresponding empirical $\beta$ values. 

The actual $p(n)$ memory functions take the analytical form Eq.\ref{eq:mf1} if we assume that the $P(E)$ distribution is a power-law function with an exponent $\gamma$. However, the fitted curves show deviations from the empirical $p(n)$ functions for small $n$ values as the measured memory functions are derived from non-perfect empirical power-law $P(E)$ distributions. Nevertheless, the analytical and empirical functions fit well asymptotically for larger $n$ values.

\end{document}